# System-Level Factors Influencing AGC Cyber-Attack Success and RAS Vulnerability in Multi-Area Grids

**Abdulmannan AJABNOOR***
**Georgia Institute of Technology**
**USA**
**aajabnoor3@gatech.edu**

**Zhi Jin ZHANG**
**The University of British Columbia**
**Canada**
**justin.zhang@ubc.ca**

**Maryam SAEEDIFARD**
**Georgia Institute of Technology**
**USA**
**maryam@ece.gatech.edu**

**Dominic GROSS**
**University of Wisconsin-Madison**
**USA**
**dominic.gross@wisc.edu**

**SUMMARY**

Automatic generation control (AGC) stabilizes frequency and scheduled power exchanges among interconnected areas, yet its reliance on communication networks introduces cyber-vulnerabilities. This work examines coordinated AGC cyber-attacks in the presence of remedial action schemes (RAS) and identifies operating conditions under which an attacker can drive grid frequency beyond safe limits while evading RAS. A feedback-based attack policy is used to manipulate AGC signals and steer system frequency toward unsafe trajectories. Then, a large nested-loop parametric sweep—spanning attacked-area inertia relative to its interconnected neighboring area, AGC capacity versus maximum tie-line transfer, AGC update rate, load damping, and bi-directional power transfers—generates over 1.45 million scenarios. These are classified using a decision tree to establish guidelines to distinguish successful from unsuccessful attacks based on frequency deviations. The resulting vulnerability guidelines are validated on a three-area system using phasor-domain simulations with IEEE 9-bus area models, detailed turbine–governor dynamics, and under/over-frequency load and generation shedding. Results show that specific combinations of AGC capacity, inertia distribution, and pre-attack tie-line loading increase the likelihood of successful attacks, offering operators actionable criteria for identifying vulnerable operating conditions.

# 1 Introduction

Power systems are critical cyber–physical infrastructures in which communication networks interface with generators and other grid assets [1]. Their reliance exposes the system to cyber-attacks capable of propagating into the physical layer and compromising stability. Physical countermeasures such as circuit breakers can help contain these events [2], but sophisticated attacks may evade detection methods and disrupt control functions [3].

This work focuses on integrity attacks targeting the communication channels of automatic generation control (AGC) and remedial action scheme (RAS). Primary frequency control relies solely on local measurements and is comparatively resilient, whereas secondary control, implemented via AGC, restores frequency and scheduled tie-line exchanges by processing remote measurements and dispatching set-points to participating units [4]. AGC uses the area control error (ACE), which combines deviations in frequency and scheduled power transfers. Because AGC relies on communicated data, tampering with its signals in one area can drive local frequency away from nominal and, through tie-line coupling, induce disturbances across interconnected neighboring areas. When primary and secondary controls cannot maintain frequency within safe bounds (e.g., ±1 Hz in Australia [5]), emergency measures such as under-frequency load shedding (UFLS), over-frequency generator shedding (OFGS), and RAS act to arrest excursions and prevent cascading failures.

Prior studies have evaluated AGC vulnerabilities to false data injection (FDI) attacks using feedback attack policy, reachability analysis, and optimal control formulations [6]–[8]. Feedback-based AGC attacks and reachability analysis have been demonstrated in a two-area setting to steer the frequency to unsafe states [6], [7]. These studies demonstrate the ability of FDI attacks to drive two or three-area systems to unsafe frequency states, but they omit protection schemes and do not characterize how system-level parameters—such as inertia, AGC capacity, tie-line capability, power-flow directions, and load damping—govern attack success in realistic multi-area settings. In contrast, [9] considers AGC cyber-attacks in the presence of RAS, but is limited to a two-area configuration, leaving open the question of how these insights extend to larger, interconnected systems with mixed or aligned power transfers.

Motivated by these gaps, this paper studies coordinated AGC cyber-attacks in multi-area power systems equipped with RAS. The attacker is assumed to compromise AGC commands, AGC- and RAS-related measurements in one targeted area (Area *A*), and phasor measurement unit (PMU) data associated with tie-line circuit breakers. A man-in-the-middle (MITM) strategy, based on the feedback attack policy in [6], is deployed to steer the frequency of Area *A* and its neighboring areas outside safe bounds. Replay attacks on RAS and AGC measurements are simultaneously used to suppress protection activation and evade operator detection until UFLS or OFGS thresholds are reached.

This study combines numerical simulations and a data-driven classification framework on three- and four-area systems to quantify how system parameters shape attack success. A nested parameter sweep varies the inertia ratio between Area *A* and its neighbors, AGC capacity of Area *A* relative to aggregate tie-line capability, AGC update rate, load damping, and pre-attack power-flow patterns, including mixed- and same-direction flows. The resulting simulation dataset is used to train a decision-tree classifier that labels attack attempts as successful or unsuccessful based on whether frequency deviations exceed safe bounds. Phasor-domain simulations on a three-area system, with each area represented by an IEEE 9-bus equivalent including turbine–governor models and UFLS/OFGS protections, validate the vulnerability guidelines and illustrate their operational relevance for identifying high-risk conditions.

## 2 Modelling of the Multi-Area Power Systems

### 2.1 Small-Signal Frequency Dynamics of a Multi-Area System

In this section, a small-signal frequency model of a multi-area power system is adopted. The attacked area $A$ is interconnected via tie-lines with neighboring areas $N_i$, $i \in \{1,2, \dots, n\}$, where $n$ denotes the number of neighboring areas, as illustrated in Fig. 1. Each area is represented by a synchronous generator (SG) equipped with primary frequency control, governor-turbine dynamics, and AGC functionality. The attacker is assumed to manipulate the AGC signal of Area $A$, denoted by $U$, while the AGC actions of all neighboring areas remain uncompromised.

The total deviation of the schedules power flow between area $j \in \{N_i, A\}$ and its neighboring areas $k \in \{N_i, A\}$ from their scheduled values is modeled following [10, Ch. 2]:

$$\frac{d}{dt}\sum_{k=1}^{n} \Delta P_{kj} = 2\pi \sum_{i=1}^{n} \hat{P}_{T_{kj}}(\Delta f_k - \Delta f_j), \tag{1}$$

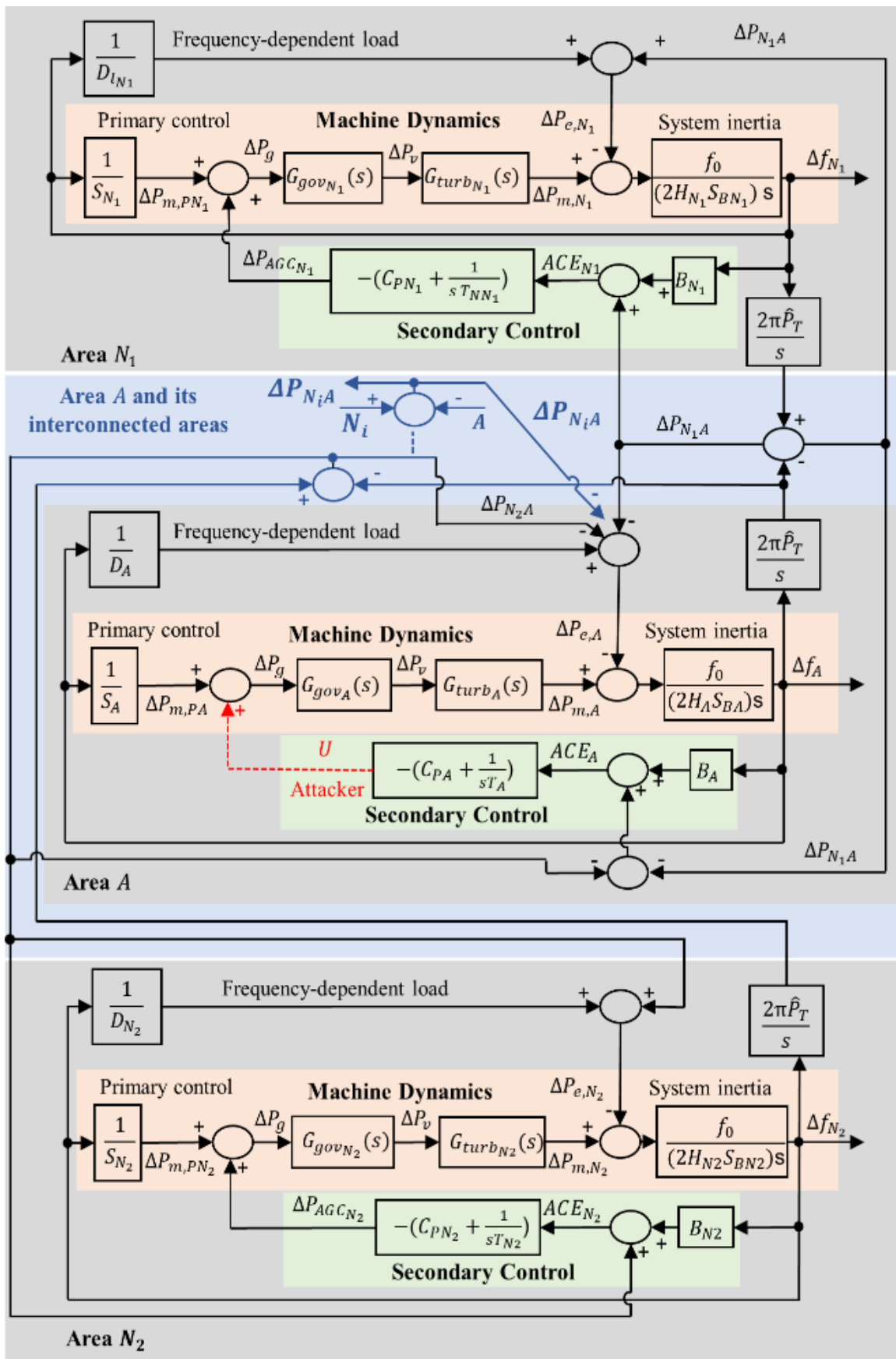

***Figure 1 – Block diagram representation of a multi-area system with machine dynamics and AGC. Areas $N_1$ and $N_2$ may also be interconnected in a meshed configuration.***

where $\hat{P}_{T_{kj}}$ is the tie-line synchronizing power coefficient, and $\Delta f_k \in \mathbb{R}$ and $\Delta f_j \in \mathbb{R}$ represent the frequency deviations of areas $k$ and $j$, respectively. The relation in (1) follows from the basic AC power flow equation $P_{kj} = P_{T_{kj}} sin(\theta_{kj})$, where $P_{kj} \in \mathbb{R}$ is the maximum transmissible power of the tie-line between areas $k$ and $j$, and $\theta_{kj} \in \mathbb{R}$ is the voltage angle difference across that line. Under a disturbance, the angle difference deviates from its pre-attack operating value, $\theta_{kj}$ is expressed as $\theta_{kj} = \theta_{0_{kj}} + \Delta\theta_{kj}$, where $\theta_{0_{kj}} \in \mathbb{R}$ denotes the steady-state pre-attack angle corresponding to the initial scheduled tie-line power flow $P_{kj}$.

The frequency deviation of area $j$ is given by [11, Ch. 2]

$$\frac{d}{dt}\Delta f_j = \frac{f_0}{2H_j S_{B_j}}\left(\Delta P_{m,j} - \Delta P_{e,j}\right), \tag{2}$$

where $f_0$ is the nominal frequency, $H_j$ is the inertia constant of area $j$, and $S_{B_j}$ is the corresponding power base. The mechanical power deviation in area $j$ consists of primary control and AGC contributions. The deviation in electrical power is a function of the total tie-line power-transfer deviation between the area $j$ and its neighboring areas $k$, along with the frequency-dependent load $D_{l_j}$:

$$\Delta P_{e,j} = \sum_{k=1}^{n} \Delta P_{kj} + \Delta f_j D_{l_j}. \tag{3}$$

The governor, $G_{\text{gov}_j}(s)$ and turbine, $G_{\text{turb}_j}(s)$ dynamics are modeled by the linear blocks illustrated in Figs. 2 and 3, respectively.

The AGC in each area is a function of its ACE, which is determined by

$$\text{ACE}_j = \sum_{k=1}^{n} \Delta P_{kj} + \frac{1}{S_j}\Delta f_j, \tag{4}$$

where $1/S_j$ is the frequency-bias factor. The AGC output deviation is then obtained from

$$\Delta P_{\text{AGC}j} = -\left(C_{p_j} + \frac{1}{sT_{N_j}}\right)\text{ACE}_j, \tag{5}$$

where $C_{p_j}$ and $T_{N_j}$ denote Proportional-Integral (PI) gains. Both primary-control and AGC set-points are subjected to saturation constraints.

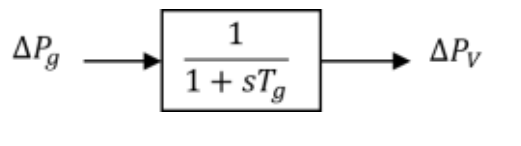

***Figure 2 - Governor linear model.***

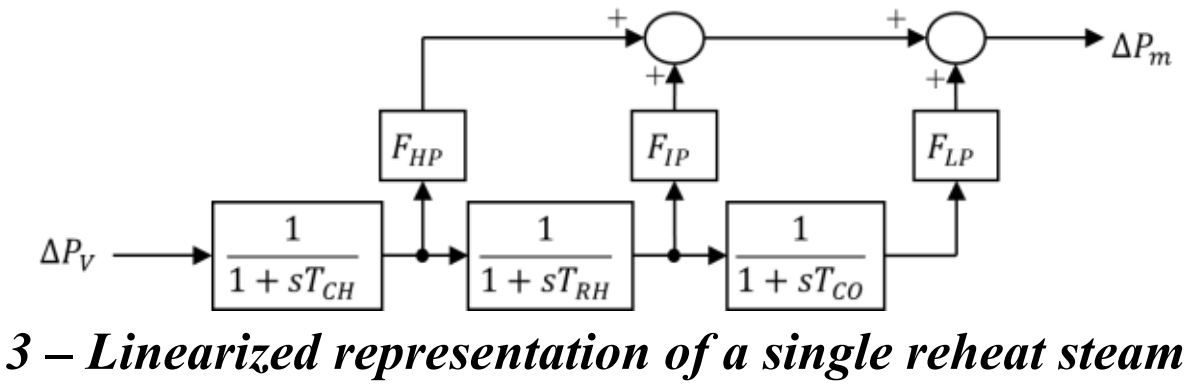

***Figure 3 – Linearized representation of a single reheat steam turbine.***

The resulting equations form a compact state-space representation in which the state vector includes area frequency deviations, tie-line angle deviations, and AGC states of the $N_i$ areas:

$$x = \left[\Delta f_j, \Delta\theta_{kj}, \Delta \mathrm{P}_{\mathrm{AGC}_{N_i}}\right]. \tag{6}$$

Use of small-signal model provides analytical insight into system stability margins, parameter sensitivities, and inter-area dynamic interactions under both normal and pre-attack conditions. Although cyberattacks may induce large disturbances, the small-signal representation is particularly valuable for: (i) deriving closed-form expressions for frequency response and inter-area oscillations to identify key vulnerability factors, (ii) characterizing system behavior near equilibrium to assess proximity to instability, and (iii) supporting decision-tree rule design by quantifying how parameter variations influence dynamics. Alternative modeling approaches—such as nonlinear time-domain models for detailed dynamics, electromechanical transient models for higher fidelity of generators and loads, and cyber-physical co-simulations are also used in cyber-attack studies. While the small-signal model does not capture all aspects of large disturbances, it complements the detailed nonlinear simulations discussed in Section 5 and provides analytical insight into system vulnerability under realistic attack scenarios.

### 2.2 Communication in Multi-Area System

In this section, communication structure in a multi-area power system is briefly reviewed. Primary frequency control relies exclusively on local measurements and thus operates without any communication link. In contrast, AGC depends on a reliable communication infrastructure to transmit measurements and dispatch control signals among areas. Prior studies have examined detailed communication architectures for a two-area system [9], in which PMU data are routed to separate phasor data concentrators (PDCs) dedicated to AGC and RAS functionalities. The communication setup adopted here extends this architecture to a three-area system, as illustrated in Fig. 4, and can be scaled up to larger configurations such as four-area systems. This structure enables coordinated exchange of frequency, tie-line power-flow, and breaker-status information required for AGC operation and RAS monitoring.

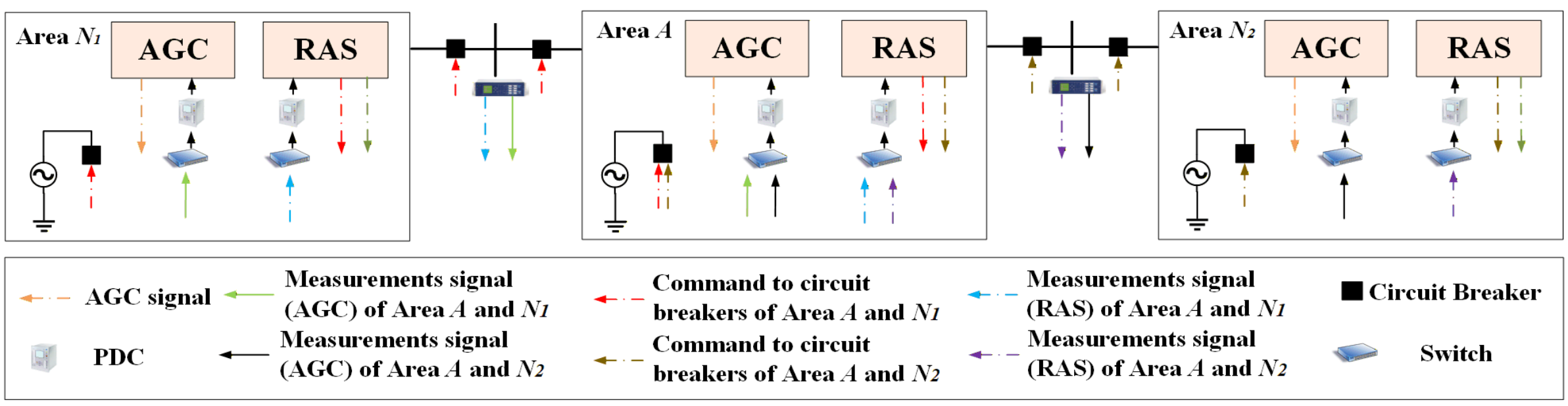

*Figure 4 - Communication configuration of the three-area system, in which both AGC and RAS have distinct PDCs.*

## 3 The Attacker Model

This section defines the attacker model, outlines feedback-based attack strategies, and develops data-driven guidelines for a multi-area system. The attacker is assumed to possess knowledge of key system parameters—including inertia, frequency bias, AGC limits, damping coefficients, and PI gains—and to be capable of compromising and manipulating AGC signals. Methods by which RAS may be bypassed are discussed in subsequent sections.

### 3.1 Review of AGC Feedback Attacks

The attack mechanism follows the feedback attack policy in [6], as illustrated in Fig. 5, where falsified AGC signals are injected to steer frequency deviations beyond safe bounds (e.g., ±1 Hz). Using $\Delta f$ and $\Delta\theta$ measurements, the attacker employs a nonlinear observer, and tunes the gain $K$ using the Metropolis–Hastings algorithm to maximize the resulting frequency deviation. The system dynamics are represented as

$$\frac{d}{dt}x = f(x, w) + g(x, w)u, \tag{7}$$

where $x = [\Delta f_j, \Delta\theta_{kj}, \Delta \mathrm{P}_{\mathrm{AGC}_{N_i}}]$ represents the state vector, $w = [\mathrm{H}_j, \theta_{0_{kj}}]$ represents the actual system parameters, and $w_0$, as shown in Fig. 5, refers to the set of parameters estimated by the attacker. If the attacker has complete knowledge of the system information, then $w_0 = w$. In this study, it is assumed that the attacker has full knowledge of system parameters.

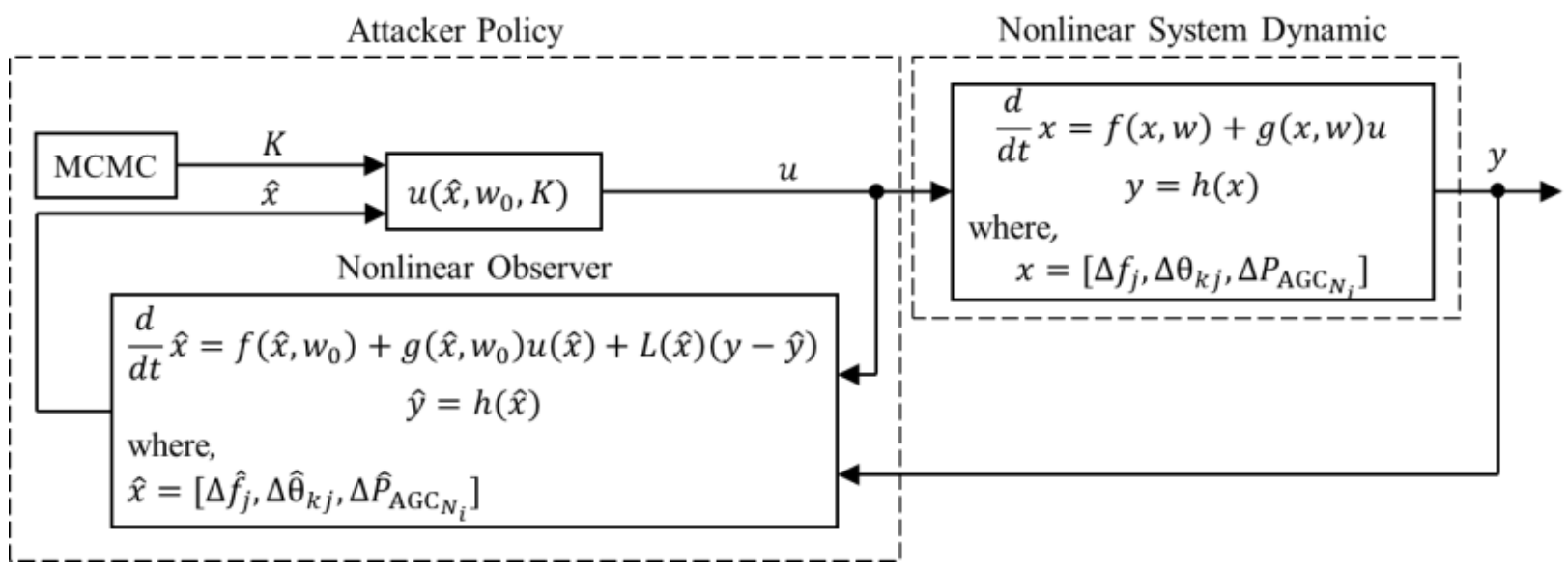

***Figure 5 - Feedback attack policy [6].***

### 3.2 Attacker Strategies

Two attack strategies may be considered for steering the system to unsafe frequency conditions from a set of $z$ possible system realizations:

1. Arbitrary Selection Strategy: The attacker randomly selects a system continuously injects falsified AGC signals. This approach increases detectability and is typically effective only when power-transfer levels exceed certain thresholds.
2. Data-Driven Strategy: The attacker uses offline simulations to establish vulnerability guidelines that identify system realizations with the highest probability of attack success. Then, the attacker estimates parameters such as inertia, maximum power-transfer capability, and AGC limits, only after a specific system realization is selected. This strategy reduces detectability and enables systematic evaluation of vulnerability across different parameter conditions.

The analysis in this paper assumes the second strategy.

### 3.3 Multi-Area Attack Guidelines and Physical Interpretation

A large set of multi-area system realizations is simulated by varying area inertias, AGC capacities, load damping coefficients, tie-line ratings, pre-attack angle differences, power flow directions, and AGC update intervals, as summarized in Table 1. Primary control bounds are scaled from the AGC limits $\Delta P_{AGC_j}^{\min}$ and $\Delta P_{AGC_j}^{\max}$. For each realization, the frequency deviation is labeled *unsafe* when $|\Delta f_j| > 1$ Hz and *safe* otherwise. A decision tree classifier is then trained to map the system parameters to these attack outcomes, yielding rule-based boundaries between successful and unsuccessful attacks. Across the simulated three- and four-area test systems, the trained decision-tree classifiers exhibit predictive performance. For the three-area system, accuracy, precision, recall, and F1-score are 87.66%, 93%, 87%, and 90%, respectively. For the four-area system, the corresponding values are 85.41%, 85%, 94%, and 89%.

The resulting decision-tree guidelines for a successful attack in a multi-area system are illustrated in Fig. 6. The determining quantities are: (i) AGC capacity of Area $A$, (ii) the total maximum tie-line transfer capability between Area $A$ and its $i$ interconnected areas $\sum_{i=1}^{n} P_{T_{N_iA}}$, and (iii) whether the aggregate neighboring inertia $\sum_{i=1}^{n} H_{N_i}$ is greater than or smaller than $H_A$. A high probability of a successful attack execution is obtained under the following conditions:

***Table 1 – Parameters used to generate data for a three-area system.***

| Parameter | Value |
|---|---|
| **Time step** | {3, 5, 6} seconds |
| $\boldsymbol{H_A}$ | {1, 2, 4, 5} |
| $\boldsymbol{H_{N_2}}$ | {1, 2, 4, 5} |
| $\boldsymbol{H_{N_1}}$ | {1, 5, 7, 9} |
| $\boldsymbol{P_{T_{N_1A}} = P_{T_{N_2A}}}$ | {225, 250, 400, 550, 700, 850, 1000, 1150, 1300, 1450} MW |
| $\boldsymbol{\theta_{0_{N_1A}}}$ | $\{-\frac{\pi}{3}, -\frac{\pi}{4}, -\frac{\pi}{6}, -\frac{\pi}{8}, \frac{\pi}{3}, \frac{\pi}{4}, \frac{\pi}{6}, \frac{\pi}{8}\}$ |
| $\boldsymbol{\theta_{0_{N_1N_2}} = \theta_{0_{N_2A}}}$ | $\{-\frac{\pi}{3}, -\frac{\pi}{4}, -\frac{\pi}{6}, -\frac{\pi}{8}, \frac{\pi}{3}, \frac{\pi}{4}, \frac{\pi}{6}, \frac{\pi}{8}\}$ |
| $\boldsymbol{\|\Delta P_{AGC}^{max}\|}$ | {350, 700, 1050, 1400} MW |
| $\boldsymbol{D_{l_j}}$ | $\{\frac{1}{0.5x200}, \frac{1}{1x200}, \frac{1}{1.5x200}\}$ Hz/MW |

1. **AGC capacity dominates tie-line limits**: A high likelihood of attack success is obtained when $\frac{AGC_A}{\sum_{i=1}^{n} P_{T_{N_iA}}} \geq 1$ because the falsified AGC signal can exceed the aggregate export/import capability of the tie-lines, regardless of inertia distribution or power-flow direction.
2. **AGC capacity below tie-line limits and interconnected neighboring area more inertial**: when $\sum_{i=1}^{n} H_{N_i} > H_A$ and $\frac{AGC_A}{\sum_{i=1}^{n} P_{T_{N_iA}}} < 1$, attack success depends on pre-attack power flows, computed as $P_{\text{pre}} = \left|P_{T_{N_iA}} \sin(\theta_{0N_iA})\right|$, and on the direction of power flow as follows:
   - **Same power-flow direction on all tie-lines** (all $\theta_{0N_iA}$ have the same sign)**:** Attack success requires

$$\left|\sum_{i=1}^{n} P_{T_{N_iA}} - \text{AGC}_A\right| < P_{\text{pre}} < \sum_{i=1}^{n} P_{T_{N_iA}}. \tag{8}$$

- **Mixed export/import directions** (some $\theta_{0N_iA} > 0$, some $\theta_{0N_iA} < 0$): The same lower bound applies as in the same power-flow direction, while the upper bound is replaced with the larger of the total export or import capability, i.e.

$$\left|\sum_{i=1}^{n} P_{T_{N_iA}} - \mathrm{AGC}_A\right| < P_{\mathrm{pre}} < \max\left(\sum_{\theta_{N_iA}>0} P_{T_{N_iA}}, \sum_{\theta_{N_iA}<0} P_{T_{N_iA}}\right). \quad (9)$$

A comparison of (8) and (9) shows that mixed power-flow directions yield a smaller upper admissible bound compared to the same-direction case.

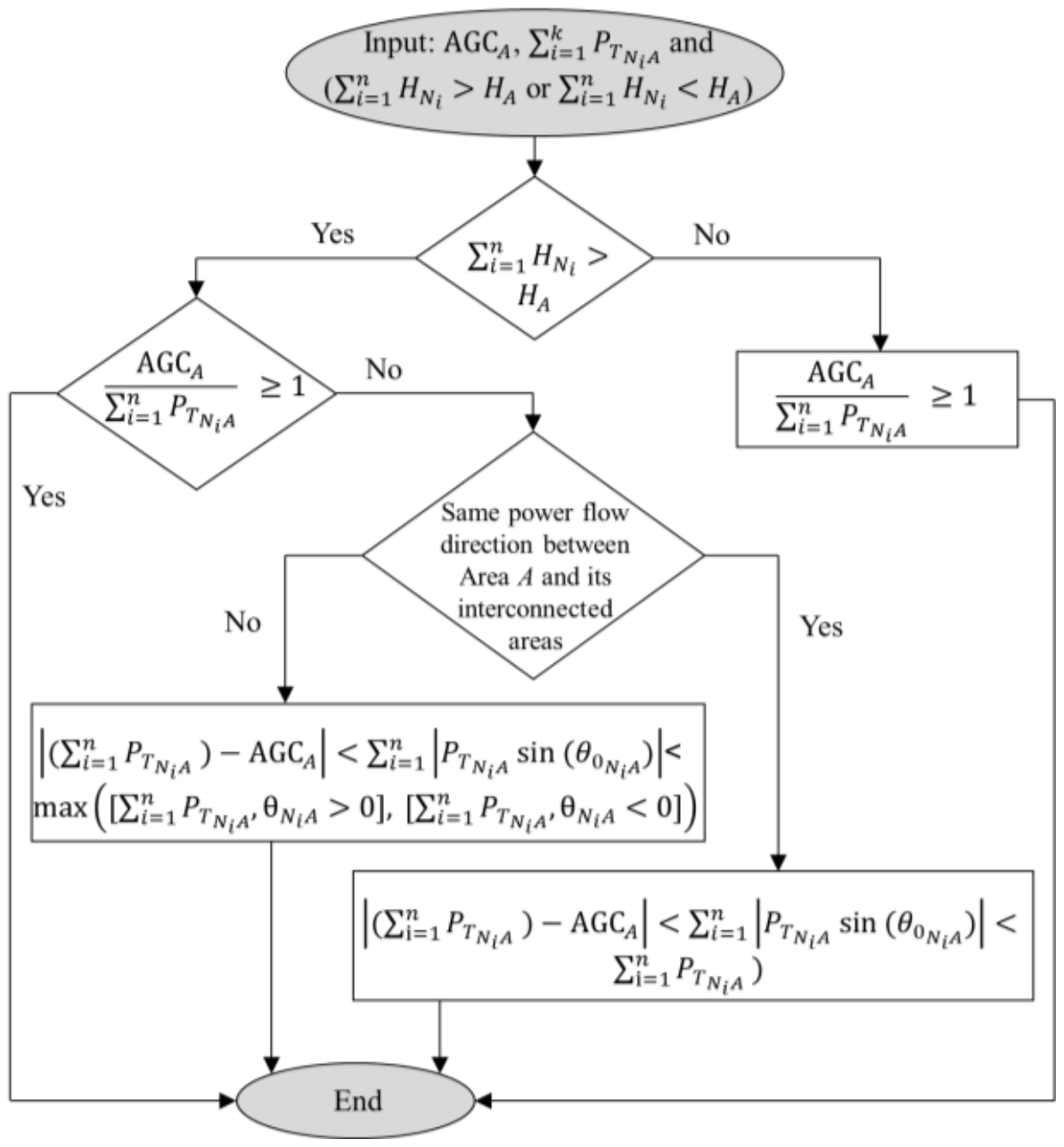

***Figure 6 - Guidelines for a multi-area system that increases the likelihood of a successful attack.***

These conditions can be interpreted through the underlying frequency dynamics. When AGC commands adjust generator outputs, high-inertia areas exhibit slower responses and greater resistance to frequency deviations, whereas low-inertia areas experience faster frequency changes for a given power imbalance. Consequently, attacks targeting an area whose inertia is lower than the aggregate inertia of its neighbors are more effective: the frequency of the targeted area can drift rapidly, while the more inertial neighboring areas respond slowly through tie-line power exchanges. Conversely, when the attacked area possesses higher inertia, its frequency is more difficult to displace, and the faster-reacting low-inertia neighbors assist in restoring balance, thereby reducing attack effectiveness.

The pre-attack angle $\theta_{0N_iA}$ indicate the loading level of each tie-line. Large values correspond to operating points near the tie-line transfer limits, leaving minimal margin for additional corrective power exchange. Under such conditions, even moderate malicious AGC injections can push tie-line flows toward or beyond their aggregate tie-line capability, producing larger frequency excursions that cannot be fully mitigated by neighboring areas. When $\mathrm{AGC}_A$ exceeds the total tie-line rating, this constraint becomes decisive, as the attacker can induce a power imbalance larger than what the network can redistribute. Following the guidelines shown in Fig. 6 yields attack success rates of 99% and 94% in three- and four-area systems, respectively.

### 3.4 System Vulnerability Guidelines to Support Cyber-Security

The proposed vulnerability guidelines can be directly used by operators in large multi-area interconnections to identify high-risk operating conditions in real time by monitoring key indicators such as area inertia, AGC capacity, and tie-line power flows. This enables targeted preventive actions—for example, committing extra generation to increase effective inertia in vulnerable areas, thereby reducing the likelihood that a coordinated AGC cyber-attack will succeed. The same indicators can also be embedded in cyber-attack detection and monitoring schemes to inform alarm weighting and tuning. Detection methods and reduced-order observers (e.g., $H_\infty$ filters in [11]) can prioritize areas flagged as vulnerable by adjusting thresholds and LMI-based gain matrices to increase sensitivity to frequency and tie-line deviations. This enhances state-estimation accuracy and supports faster detection of false-data injection attacks.

### 3.5 Attack Applicability to Other Grid Topologies

While this study focuses on coordinated AGC and RAS attacks in a multi-area system, the proposed framework can be extended to other grid topologies, including meshed networks. Since the analysis is rooted in system-level parameters, the guidelines remain applicable to meshed AC networks, particularly when SGs constitute the primary generation sources. In such cases, the vulnerability of the attacked area and its immediate neighbors predominantly govern the dynamic response, and the guidelines derived here are expected to hold.

In systems with substantial penetration of power electronic interfaces—such as HVDC links or inverter-based resources (IBRs)—the frequency dynamics may differ significantly from SG-dominated systems. In these settings, converter control limits, reduced inertia, and altered interaction dynamics can modify the attack surface. Consequently, although the methodology can still be applied, the vulnerability-assessment guidelines may require refinement to account for the distinct behaviors introduced by power-electronic resources.

## 4 Impact of Communication and Protection

The primary objective of the attacker is to evade RAS actions long enough to drive the system into UFLS or OFGS operation. UFLS sheds load in stages when frequency drops below set thresholds, while OFGS disconnects generators if frequency exceeds safe limits, both intended to protect equipment and preserve stability. RAS schemes monitor abnormal operating conditions (e.g., out-of-step behavior or angular instability) and may trip tie-line circuit breakers or generators to isolate affected areas, thereby blocking the progression of an attack. Manual operator intervention, initiated in response to abnormal indicators on the human–machine interface (HMI), can further decrease the likelihood of a successful cyber intrusion.

In a coordinated multi-area attack, as illustrated in Fig. 7, the adversary simultaneously manipulates the AGC signal of the targeted area while replaying falsified measurements to both AGC and RAS. By corrupting PMU data observed by RAS, the attacker can block tripping of tie-line and generator breakers, thereby maintaining interconnection among areas and enabling continuous malicious AGC injections. Under these conditions, adherence to the previously derived vulnerability guidelines is required to drive the system toward UFLS or OFGS activation while avoiding early detection and protective isolation.

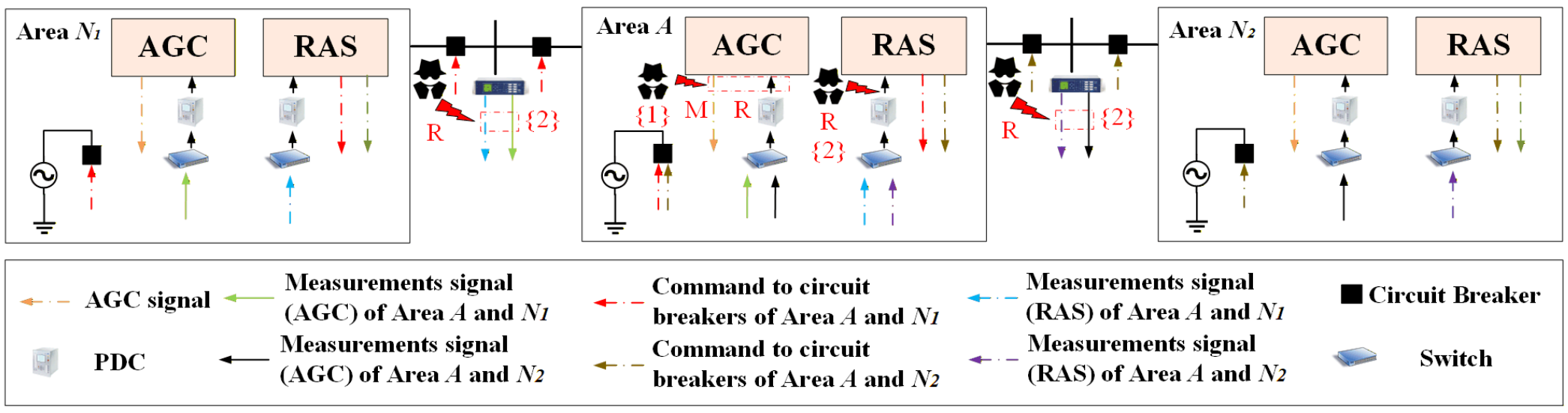

{1} The attacker compromises both the AGC and measurements signal sent to the AGC of the attacked area (Area $A$). {2} Simultaneously, the attacker compromises RAS measurement of Area $A$ and the tie-lines measurement signals to avoid the opening of circuit breakers on tie-lines and generators.

***Figure 7 - Detailed scenario of a coordinated three-area attack to derive areas A, $N_1$, and $N_2$ to instability.***

## 5 Numerical Results

This section focuses on coordinated attacks that satisfy the derived guidelines. An attacker is assumed to target Area $A$ in a three-area system, either System 1 or System 2. The systems differ in total tie-line capacity, AGC limits, power-flow directions, and inertia ordering between Area $A$ and its neighboring areas, as shown in Table 2. According to the guidelines, System 1 is vulnerable for any pre-attack angle because $\frac{AGC_A}{P_{T_{N1A}}+P_{T_{N2A}}} > 1$. In System 2, the conditions in (8) indicate that successful attacks require $100 < P_{pre} < 500$, while in System 3, using (9), successful attacks require $100 < P_{pre} < 250$. Thus, System 1 is the more attractive target.

***Table 2 – Three independent system options available to the attacker.***

| System | $P_{T_{N_1A}} + P_{T_{N_2A}}$ (MW) | Max $AGC_A$ (MW) | Inertia | Power flow |
|---|---|---|---|---|
| **1** | 500 | 750 | $H_{N1} + H_{N2} < H_A$ | Same direction |
| **2** | 500 | 400 | $H_{N1} + H_{N2} > H_A$ | Same direction |
| **3** | 250 ($\theta_{N1A} > 0$)<br>250 ($\theta_{N2A} < 0$) | 400 | $H_{N1} + H_{N2} > H_A$ | Different direction |

Each area is modeled as an IEEE 9-bus system with three identical generators (equipped with turbine–governor–exciter models) and three loads, interconnected by a tie-line and simulated in a phasor-domain environment. Frequencies at all generator and load buses are monitored, and circuit breakers trip on over-frequency for generators and under-frequency for loads. To emulate staged protection, a two-second delay is enforced between successive tripping events within each area. A replay attack is used to mask out-of-step or angular instability indicators from the protection system, thereby preventing tie-line breaker operations.

For System 1, the attacker estimates $\sum_{i=1}^{2} H_{Ni} = 4$, $H_A = 6$ ($\sum_{i=1}^{2} H_{Ni} < H_A$) and selects $P_{pre} = 250$ MW. Each area generates 3000 MW, with a total load of 2000 MW in Area $A$ and 2250 MW in other areas. As shown in Fig. 8, once the attack begins at $t = 120\,s$, frequencies in all areas diverge and staged tripping ultimately produces a blackout. In System 2, the attacker assumes $\sum_{i=1}^{2} H_{Ni} = 4$, $H_A = 2$ ($\sum_{i=1}^{2} H_{Ni} > H_A$) with $P_{pre} > 100$ MW. Total generation remains 3000 MW per area, while loads are 2250 MW in Areas $N_1 and\ N_2$ and 2350 MW in Area $A$. Figure 9 confirms that the attack again drives all three areas to blackout. System 3 exhibits similar behavior when $P_{pre} > 100$, as shown in Fig. 10. In contrast, when $P_{pre} < 100$ MW, the same attack fails to destabilize the system, consistent with the decision-tree-based guidelines; Fig. 11 shows the corresponding stable outcome.

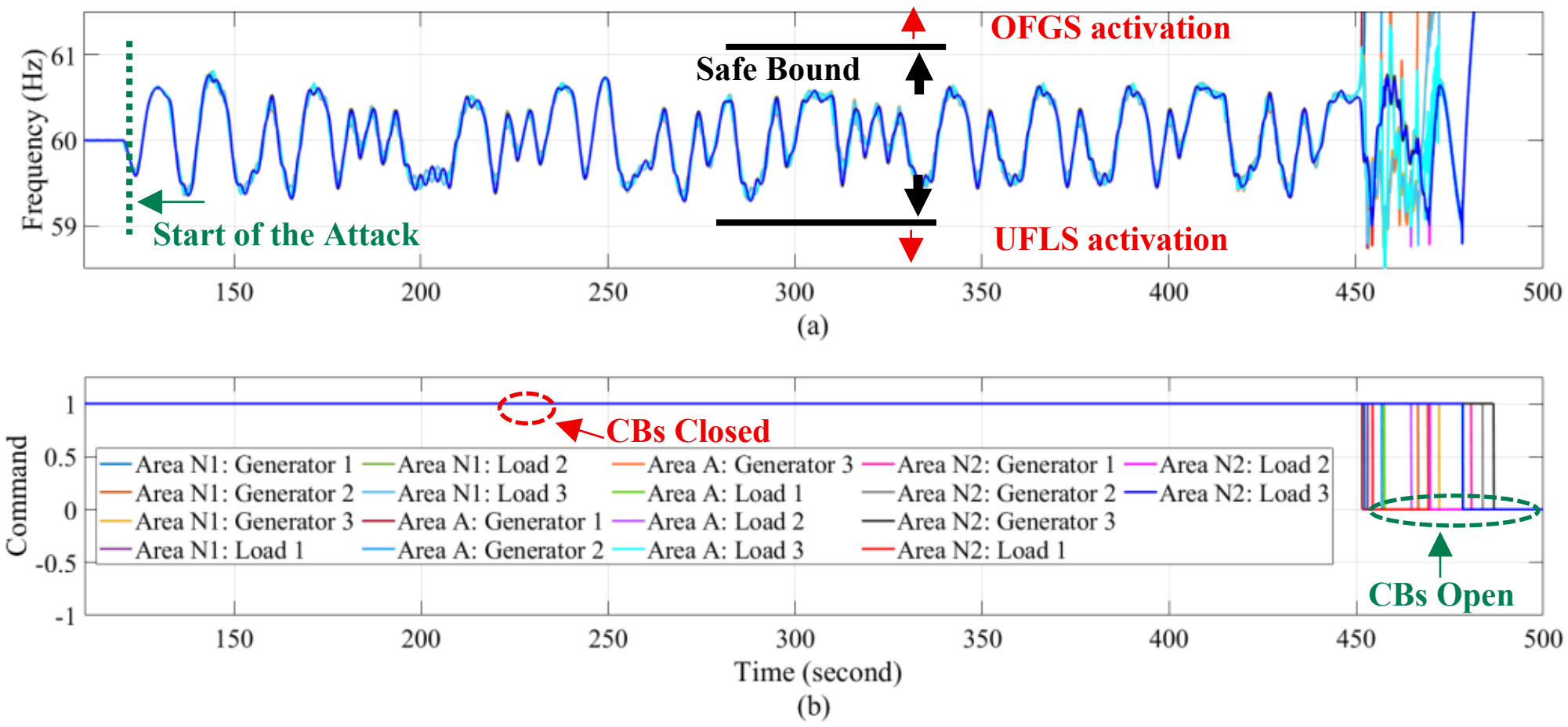

***Figure 8 - Generators and loads of System 1 with same-direction transfers and $AGC_A$ exceeding total tie-lines capacity: (a) frequencies, and (b) circuit breaker commands.***

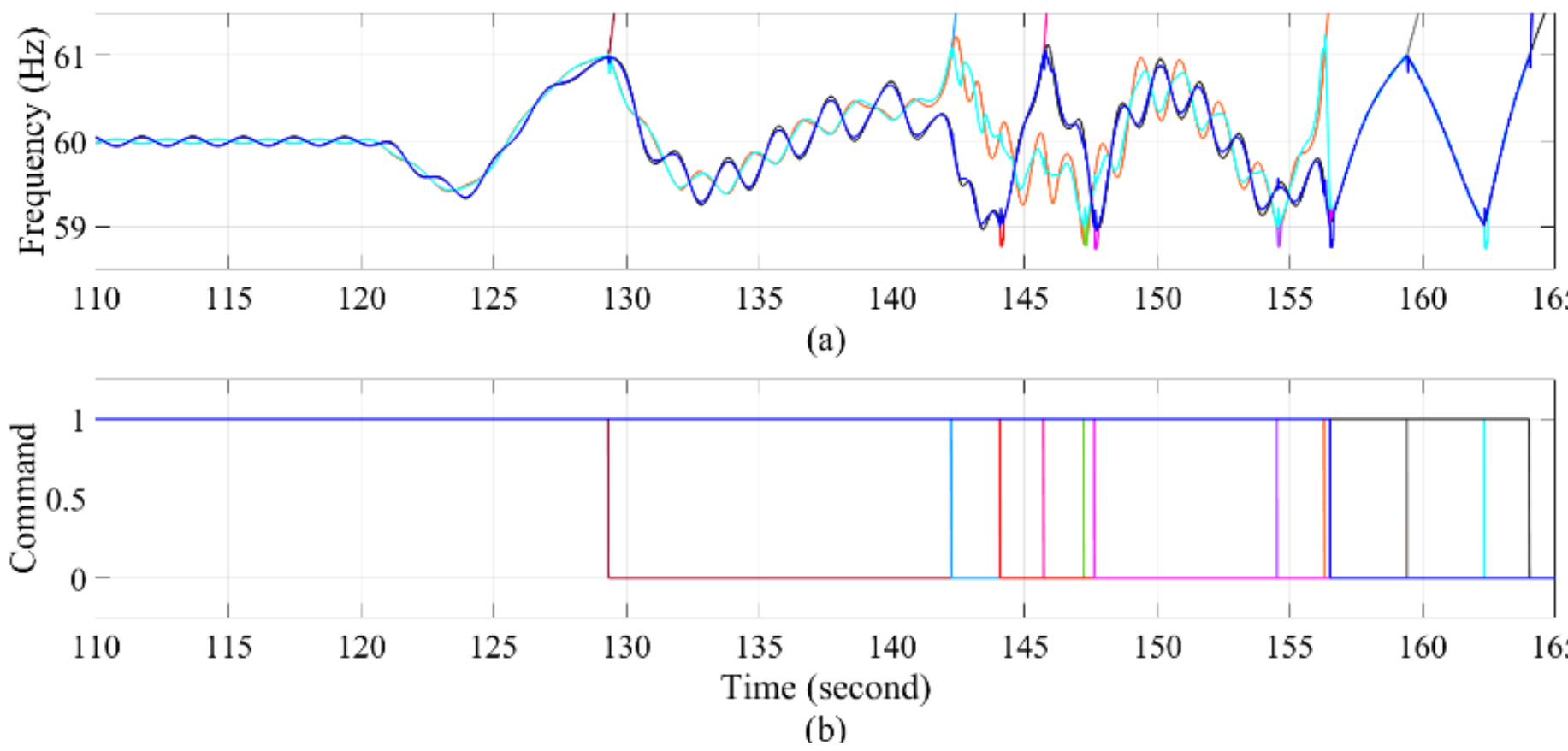

***Figure 9 - Generators and loads of System 2 with same-direction transfers and $P_{pre}$ above the threshold: (a) frequencies, and (b) circuit breaker commands. The color-coding follows Fig. 8.***

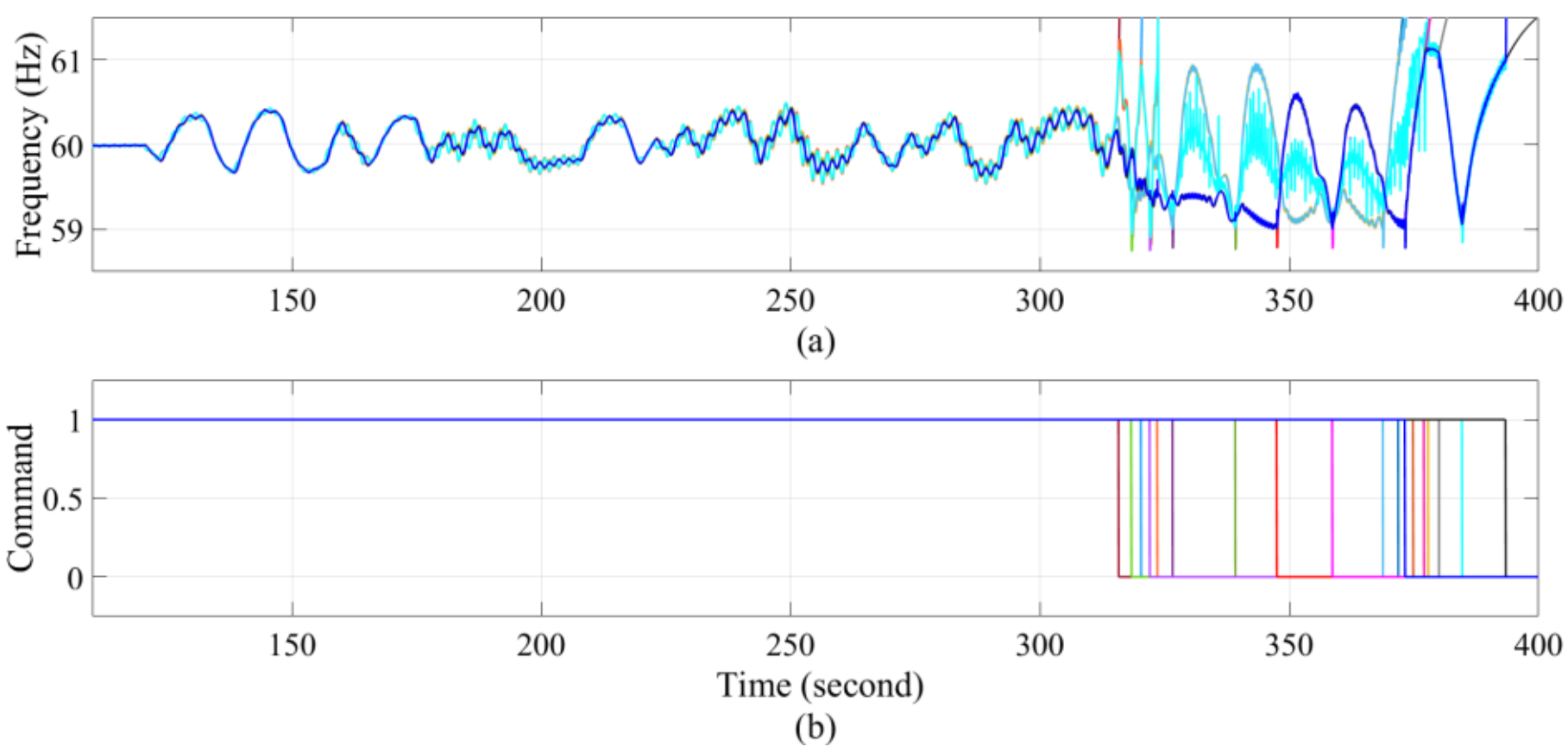

***Figure 10 - Generators and loads of System 3 with different direction transfers and $P_{pre}$ above the threshold: (a) frequencies, and (b) circuit breaker commands. The color-coding follows Fig. 8.***

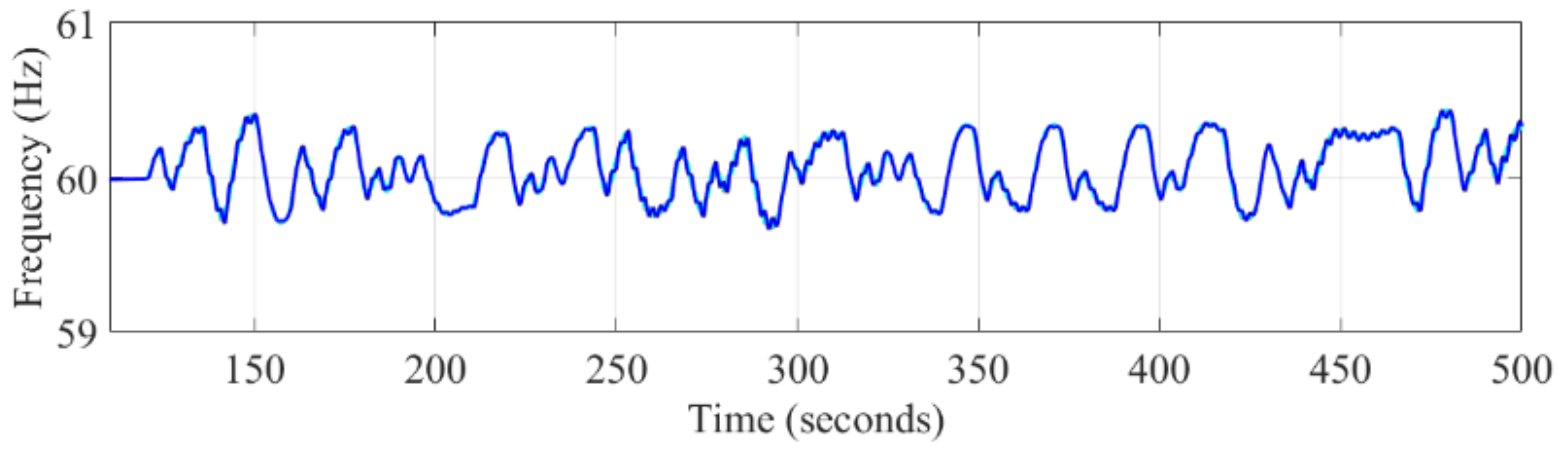

***Figure 11 - Generators and loads of System 2 when guidelines are not satisfied: (a) frequencies, and (b) circuit breaker commands.***

# 6 Conclusions

This work analyzed how key system parameters in multi-area systems influence the success of AGC cyber-attacks under an attacker model with access to both AGC signals and AGC/RAS measurements. A decision-tree classifier was trained on a large set of simulated system realizations to identify worst-case operating conditions and derive vulnerability guidelines. Phasor-domain simulations of a protected multi-area IEEE 9-bus system validated these guidelines and demonstrated how parameter-dependent attacks can drive the system toward UFLS or OFGS activation thresholds.